\documentclass[aps,prl,reprint,superscriptaddress]{revtex4-2}
\usepackage{psfrag,slashed,cancel,array,graphicx,todonotes,hyperref}
\usepackage[utf8]{inputenc}
\usepackage{mathtools}
\usepackage{mathrsfs}
\usepackage{multirow}
\usepackage{braket}
\usepackage{slashed}
\usepackage{booktabs} 
\usepackage[normalem]{ulem}
\usepackage{slashed}
\hypersetup{pdftitle={},pdfcreator={},linkcolor=[rgb]{0.15,0.35,0.75},colorlinks=true,citecolor=[rgb]{0.675,0,0.2},urlcolor=[rgb]{0.15,0.35,0.65}}

\def\beq{\begin{equation}}
\def\eeq{\end{equation}}
\def\bsp#1\esp{\begin{split}#1\end{split}}
\def\d{{\rm d}}

\newcommand{\nn}{\nonumber}

\AtBeginDocument{
\heavyrulewidth=.08em
\lightrulewidth=.05em
\cmidrulewidth=.03em
\belowrulesep=.65ex
\belowbottomsep=0pt
\aboverulesep=.4ex
\abovetopsep=0pt
\cmidrulesep=\doublerulesep
\cmidrulekern=.5em
\defaultaddspace=.5em
}

\def\be{\begin{equation}}
\def\ee{\end{equation}}

\begin{document}
\preprint{CERN-TH-2026-185}

\author{Liang Dong}
\email{liang.dong@sjtu.edu.cn}
\affiliation{State Key Laboratory of Dark Matter Physics, Shanghai Key Laboratory for Particle Physics and Cosmology, Key Laboratory for Particle Astrophysics and Cosmology (MOE),
School of Physics and Astronomy, Shanghai Jiao Tong University, Shanghai 200240, China}
\author{Shen Fang}
\email{sfang23@m.fudan.edu.cn}
\affiliation{Department of Physics, Center for Field Theory and Particle Physics, and Key Laboratory of Nuclear Physics and Ion-beam Application (MOE), Fudan University, Shanghai, 200433, China}
\author{Jun Gao}
\email{jung49@sjtu.edu.cn}
\affiliation{State Key Laboratory of Dark Matter Physics, Shanghai Key Laboratory for Particle Physics and Cosmology, Key Laboratory for Particle Astrophysics and Cosmology (MOE),
School of Physics and Astronomy, Shanghai Jiao Tong University, Shanghai 200240, China}
\affiliation{Southern Center for Nuclear-Science Theory (SCNT), Institute of Modern Physics, Chinese Academy of Sciences, Huizhou 516000, Guangdong Province, China}
\author{Hai Tao Li}
\email{haitao.li@sdu.edu.cn}
\affiliation{School of Physics, Shandong University, Jinan, Shandong 250100, China}
\author{Ding Yu Shao}
\email{dyshao@fudan.edu.cn}
\affiliation{Department of Physics, Center for Field Theory and Particle Physics, and Key Laboratory of Nuclear Physics and Ion-beam Application (MOE), Fudan University, Shanghai, 200433, China}
\affiliation{Southern Center for Nuclear-Science Theory (SCNT), Institute of Modern Physics, Chinese Academy of Sciences, Huizhou 516000, Guangdong Province, China}
\affiliation{Shanghai Research Center for Theoretical Nuclear Physics, NSFC and Fudan University, Shanghai 200438, China}
\affiliation{Center for High Energy Physics, Peking University, Beijing 100871, China}
\author{Bin Zhou}
\email{zhoubin@hainanu.edu.cn}
\affiliation{Center for Theoretical Physics, Hainan University, Haikou 570228, China}
\affiliation{School of Physics and Optoelectronic Engineering, Hainan University,
Haikou, 570228, China}
\author{Yu Jiao Zhu}
\email{yu.jiao.zhu@cern.ch}
\affiliation{CERN, Theoretical Physics Department, CH-1211 Geneva 23, Switzerland}

\title{Polarized Semi-Inclusive Deep-Inelastic Scattering at $\mathcal O(\alpha_s^3)$ in QCD}

\begin{abstract}
Unraveling the partonic origin of the proton spin requires precise determinations of polarized parton distribution functions (PDFs), which depend on comparably precise theoretical predictions for polarized scattering, particularly in view of the high-precision measurements anticipated at the future Electron-Ion Collider.
We present the first next-to-next-to-next-to-leading order (N$^3$LO) QCD predictions for longitudinally polarized semi-inclusive deep-inelastic scattering (SIDIS) in a fully differential form, together with next-to-next-to-leading order predictions for the hadron transverse-momentum spectrum.
These results are obtained by extending the two-dimensional transverse-momentum subtraction framework to the spin-dependent cross section.
Together with the corresponding unpolarized calculation, these results enable consistent N$^3$LO predictions for longitudinal double-spin asymmetries and provide a precision baseline for future analyses of helicity PDFs and transverse-momentum-dependent helicity distributions.
\end{abstract}

\maketitle

\paragraph*{Introduction.---}
Understanding the partonic origin of the proton spin remains a fundamental challenge in quantum chromodynamics (QCD). 
The European Muon Collaboration revealed that quark helicities account for only a limited fraction of the proton spin~\cite{EuropeanMuon:1987isl, EuropeanMuon:1989yki}. 
Precise determinations of polarized parton distribution functions (pPDFs) are therefore essential for quantifying the quark and gluon helicity contributions to the proton spin.
Over the past decades, global QCD analyses of helicity PDFs have combined measurements from polarized deep-inelastic scattering (DIS), semi-inclusive DIS (SIDIS), and high-energy polarized proton-proton collisions~\cite{Borsa:2024mss, Bertone:2024taw, Cocuzza:2025qvf, Cruz-Martinez:2025ahf}. 
In addition, lattice QCD provides complementary constraints, particularly on the sea-quark flavor asymmetry~\cite{Lin:2017snn, Constantinou:2020hdm, Chen:2016utp, Bringewatt:2020ixn, Holligan:2024wpv, LatticeParton:2025eui, zhao2026totalgluonhelicitycontribution, Gao:2026wlz}.

By identifying final-state hadrons, longitudinally polarized SIDIS provides the flavor sensitivity needed to disentangle quark and antiquark helicity distributions.
The precision and kinematic reach anticipated at the forthcoming Electron-Ion Collider (EIC)~\cite{AbdulKhalek:2021gbh, AbdulKhalek:2022hcn} demand higher-order predictions for polarized SIDIS that implement realistic experimental cuts.
Several ingredients required for precision polarized-SIDIS predictions have recently become available at high perturbative orders.
The DGLAP evolution kernels for polarized PDFs are now known through next-to-next-to-leading order (NNLO)~\cite{Moch:2014sna, Blumlein:2021enk, Blumlein:2021ryt, Blumlein:2022gpp, Zhu:2025gts, Behring:2025avs}, alongside the Wilson coefficients for polarized DIS structure functions~\cite{Blumlein:2022gpp, Borsa:2022irn}. 
To enable fully differential predictions, the infrared behavior of polarized NNLO matrix elements has been comprehensively analyzed~\cite{Gehrmann:2025xab}, facilitating the construction of appropriate subtraction terms~\cite{Gehrmann-DeRidder:2005btv}. 
Phenomenologically, many efforts have been devoted to these applications, yielding NLO cross sections for dijet production~\cite{Borsa:2021afb} and the NNLO differential cross sections for inclusive jet production in polarized DIS~\cite{Borsa:2020ulb, Borsa:2020yxh}. 
Following the NLO calculation~\cite{deFlorian:1997zj} and approximate NNLO results~\cite{Abele:2021nyo}, complete NNLO coefficient functions have recently been obtained for polarized SIDIS~\cite{Bonino:2024wgg, Goyal:2024tmo,Goyal:2024emo, Bonino:2025bqa}, together with mixed QCD$\otimes$QED corrections~\cite{Goyal:2025qyu}.
Extending fully differential polarized-SIDIS predictions to N$^3$LO requires dedicated control of increasingly intricate infrared singularities.
The conventional $q_T$ subtraction method~\cite{Catani:2007vq, Catani:2009sm, Catani:2010en, Catani:2011qz} has recently been extended to identified hadron production~\cite{Fu:2024fgj, Gao:2026tnd}. 
Building on this development, a two-dimensional transverse-momentum subtraction framework was introduced and applied to unpolarized SIDIS at N$^3$LO~\cite{Dong:2026eas, Dong:2026uvw}.
The framework partitions phase space into two unresolved regions governed by complementary transverse-momentum-dependent (TMD) factorization theorems and a fully resolved region evaluated with fixed-order matrix elements.
In this Letter, we extend the two-dimensional subtraction framework to longitudinally polarized SIDIS and present the first fully differential N$^3$LO QCD predictions for this process. 
In addition, we provide the first NNLO predictions for hadron production at finite transverse momentum in polarized SIDIS. 
At EIC kinematics, the N$^3$LO corrections reach approximately $10\%$ at large $x$ and substantially reduce the residual scale dependence, while the NNLO corrections to the finite-$P_{hT}$ spectrum are $13$--$20\%$ over the range studied.
Together with the corresponding unpolarized calculation, these results enable consistent N$^3$LO predictions for longitudinal spin asymmetries in SIDIS and provide a precision baseline for future extractions of helicity PDFs at the EIC.

\begin{figure*}[t]
    \centering
    \includegraphics[width=0.95\textwidth]{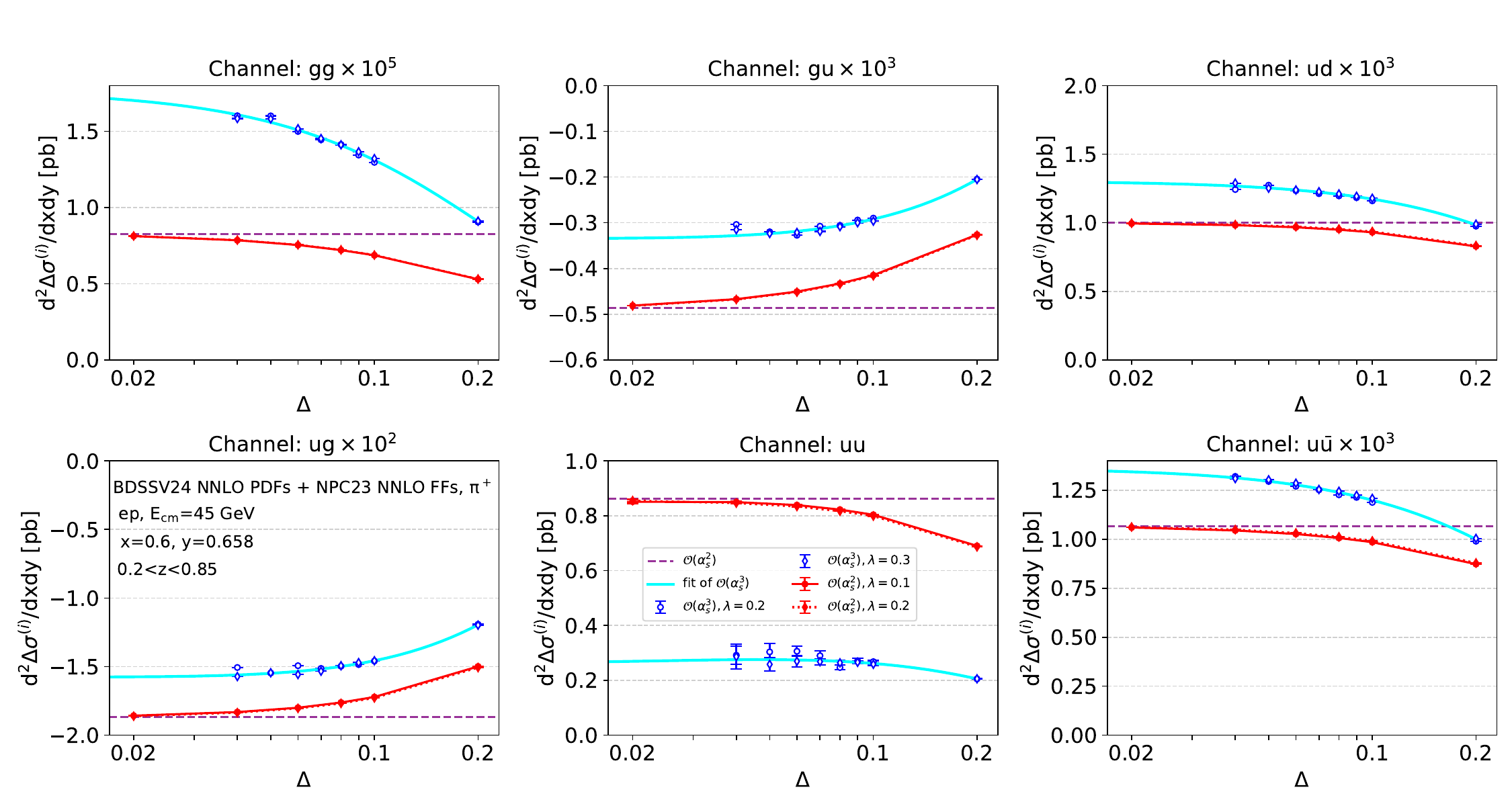}
    \caption{
    Cutoff dependence of the spin-dependent cross sections for inclusive $\pi^+$ production in polarized $ep$ collisions at the center-of-mass energy $E_{\rm cm}=45$~GeV, with $x=0.6$, $y=0.658$, and $0.2<z<0.85$. 
    The panels display the $\mathcal{O}(\alpha_s^2)$ (red) and $\mathcal{O}(\alpha_s^3)$ (blue) contributions for six representative partonic channels. 
    Error bars indicate Monte Carlo statistical uncertainties.
    The solid cyan lines represent a fit to the asymptotic limit $\Delta \to 0$. 
    The dashed purple lines denote the reference values of $\mathcal{O}(\alpha_s^2)$.
    }
    \label{fig:cutoff_dep}
\end{figure*}

\paragraph*{The method.---}
We extend the two-dimensional transverse-momentum subtraction method of Ref.~\cite{Dong:2026uvw} to the spin-dependent cross section $\Delta\sigma\equiv \tfrac12(\sigma^{++}-\sigma^{+-})$, where the superscripts label the relative orientations of the lepton and nucleon helicities.
The kinematics and phase-space partition follow the unpolarized construction~\cite{Dong:2026uvw}.
In the Breit frame, we define $P_{hT}=P_h\sin\delta\theta$ and $p_{\rm out}=P_{hT}\sin\delta\phi$, where $\delta\phi$ is measured relative to the photon--jet plane using the jet definition of Ref.~\cite{Dong:2026uvw}.
The limits $\delta\theta\to0$ and $\delta\phi\to0$ define two complementary unresolved regions governed by TMD factorization~\cite{Gao:2026tnd, Dong:2026eas, Fu:2024fgj}; configurations away from both limits form the resolved region.
Introducing $\delta\theta^{\rm cut}=\Delta$ and $\delta\phi^{\rm cut}=\lambda\Delta$, we write
\begin{align}
     \frac{\d\Delta\sigma}{\d\mathcal{O}}  & = \! \underbrace{\int_{0}^{\Delta} \d \delta\theta \frac{\d\Delta\sigma^A}{ \d\delta\theta {\d\mathcal{O}} }}_{A} + \underbrace{\int_{\Delta}^{\delta\theta ^{\rm max}} \d \delta\theta  \int_{0}^{\lambda \Delta} \d \delta\phi \frac{\d\Delta\sigma^{B}}{ \d \delta\theta \d\delta\phi{\d\mathcal{O}} }}_{B} \notag\\ & 
      +\underbrace{ \int_{\Delta}^{\delta\theta ^{\rm max}} \d \delta\theta \int_{\lambda \Delta}^{\delta\phi^{\rm max}} \d \delta\phi \frac{\d\Delta\sigma^{C}}{ \d \delta\theta \d\delta\phi{\d\mathcal{O}} }}_{C}\,,
\end{align}
where regions $A$ and $B$ are evaluated from their leading-power TMD factorization theorems, whereas region $C$ is obtained from fixed-order matrix elements.
Power corrections are suppressed by taking $\Delta,\lambda\ll1$.
For region~$A$, the spin-dependent cross section factorizes as
\begin{align} \label{eq:sidis}
      &\frac{\d\Delta\sigma^A}{\d x \, \d y \, \d z \, \d^2 \vec P_{hT} }  \propto \int \frac{\d^2 \vec{b}_{\perp}}{4\pi^2} e^{-i \vec P_{hT} \cdot \vec{b}_\perp/z}  \sum_{i}  \Delta H_{ei\to ei}(Q)
      \nn\\
  &\hspace{0.5cm}\times  \Delta{\cal B}_{i/p}(x,\vec{b}_\perp)\,
      {\cal D}_{h/i}(z,\vec{b}_\perp) \, S_{qq}(\vec{b}_\perp )\left[1 + \mathcal{O}(\Delta )\right] 
\end{align}
up to power corrections that vanish as $\Delta\to0$.
Here $Q$, $x$, $y$, and $z$ are the momentum transfer, Bjorken variable, inelasticity and identified hadron momentum fraction, respectively.
The polarized hard function $\Delta H_{ei\to ei}$, which encodes the difference of squared helicity amplitudes for the $\gamma^*q\to q$ process, is extracted from the three-loop polarized quark form factor~\cite{Becher:2006mr, Moch:2005tm, Moch:2005id, Baikov:2009bg, Gehrmann:2010ue, Gehrmann:2010tu, Lee:2022nhh}.
The helicity TMD beam function $\Delta{\cal B}_{i/p}$ is matched onto the helicity PDFs $\Delta f_{a/p}$ via
\begin{align}
  \Delta{\cal B}_{i/p}(x,\vec{b}_\perp) = \sum_a \int_x^1 \frac{\d{\hat x}}{\hat x} \,
  \Delta{\cal I}_{i/a}(\hat x,\vec{b}_\perp) \,
  \Delta f_{a/p}\!\left(\frac{x}{\hat x}, \mu\right)\,,
\end{align}
with $\Delta{\cal I}_{i/a}$ denoting the perturbative matching coefficients that relate the TMD parton distribution to the corresponding collinear PDF. 
The N$^3$LO coefficient functions $\Delta{\cal I}_{i/a}$ are available in Ref.~\cite{Zhu:2025gts} in the $\overline{\mathrm{MS}}$ scheme, except for the three-loop quark flavor-changing piece $\Delta{\cal I}^{(3)}_{q/q'}$. 
We use the corresponding result in the \textsc{HVBM} scheme~\cite{tHooft:1972tcz,Breitenlohner:1977hr} for its finite part, while the $q_T$-dependent contributions in $\overline{\mathrm{MS}}$ are determined by renormalization-group evolution, resulting in the cutoff independence observed in the quark flavor-changing channel in Fig.~\ref{fig:cutoff_dep}.
Moreover, this scheme difference vanishes in the threshold limit $\hat{x}\to1$~\cite{Zhu:2025gts}.
The TMD fragmentation function ${\cal D}_{h/i}$ and soft function $S_{qq}$ remain identical to the unpolarized case~\cite{Luo:2020epw,Li:2016ctv,Zhu:2020ftr}, since neither the fragmentation nor soft radiation is sensitive to the initial-state helicity.
The TMD factorization formula for the spin-dependent cross section in region $B$ reads
\begin{align} \label{eq:fact_phi_jet}
      &\frac{\d\Delta \sigma^B}{\d x \, \d y \, \d z \, \d^2 \vec P_{hT}  \, \d p_{\rm out}} \propto
      \int \frac{\d b}{2\pi}e^{i p_{\rm out} b/\zeta} \\
  &\hspace{1cm}\times   \sum_{ijk} \int \d \xi\,   \Delta H_{ei\to ejk}(Q,\xi,\zeta)\nn\\
  &\hspace{1cm}\times  \Delta {\cal B}_{i/p}(\xi,b)\,
      {\cal D}_{h/j}(\zeta,b)\,{\cal J}_{k}(b)  \, S_{ij k}(b)\left[1 + \mathcal{O}(\lambda)\right] \nn
\end{align}
at leading power, which is obtained from its unpolarized counterpart~\cite{Dong:2026eas} by replacing the unpolarized hard function and TMD beam function with their polarized counterparts.
The spin-dependent cross sections can be calculated up to power corrections that vanish in the limit $\lambda\to0$. 
Although $h_{1L}^{\perp g}$, describing linearly polarized gluons in a longitudinally polarized nucleon, is a leading-twist gluon TMD, its contribution to the perturbative subtraction term is suppressed by $\Lambda_{\rm QCD}/q_T$ and is therefore beyond the leading-power accuracy retained here~\cite{Boussarie:2023izj, Lyubovitskij:2021qza}.
This differs from the unpolarized case, where linearly polarized
gluon TMDs contribute.
The hard functions $\Delta H_{ei\to ejk}$ are required through two loops and are obtained from Refs.~\cite{Garland:2002ak, Gehrmann:2009vu}, in complete agreement with the results of Refs.~\cite{Gehrmann:2022vuk, Gehrmann:2023zpz, NNLOJET:2025rno}.
The TMD jet functions $\mathcal{J}_k$, including the linearly polarized gluon jet function~\cite{Gutierrez-Reyes:2019vbx, Fang:2024auf, Bell:2021dpb, Brune:2022cgr, Buonocore:2025ysd}, and the soft function $S_{ijk}(b)$~\cite{Echevarria:2015byo, Lubbert:2016rku,Gao:2019ojf, Chien:2020hzh, Chien:2022wiq, Gao:2023ivm, Fu:2024fgj}, coincide with those in the unpolarized case.
For a detailed discussion of the kinematic variables ($\xi$ and $\zeta$), and the hard, soft, and jet functions, we refer the reader to Ref.~\cite{Dong:2026eas}.
For the resolved region~$C$, the spin-dependent cross section corresponds to the production of hadron plus dijet, which can be calculated within the \texttt{FMNLO} framework for identified hadron production~\cite{Liu:2023fsq}.
The extension of \texttt{FMNLO} to polarized SIDIS calculations relies on the dipole subtraction scheme for polarized initial hadrons~\cite{Catani:1996jh,Borsa:2020yxh}. 
We extract and reconstruct the required polarized amplitudes from the helicity amplitudes provided in Refs.~\cite{Bern:1997sc, Campbell:2002tg, Campbell:2003hd, Campbell:2010ff}.

Upon combining regions $A$, $B$, and $C$, the leading-power dependence on $\Delta$ and $\lambda$ cancels, while the residual power corrections vanish as $\Delta,\lambda\to0$.
With all ingredients available through $\mathcal O(\alpha_s^3)$, this limit yields the N$^3$LO polarized-SIDIS cross section.
At finite $P_{hT}$, only regions $B$ and $C$ contribute, providing fully differential NNLO predictions relevant to TMD-PDF extractions.
\begin{figure}[t]
    \includegraphics[width=0.95\linewidth]{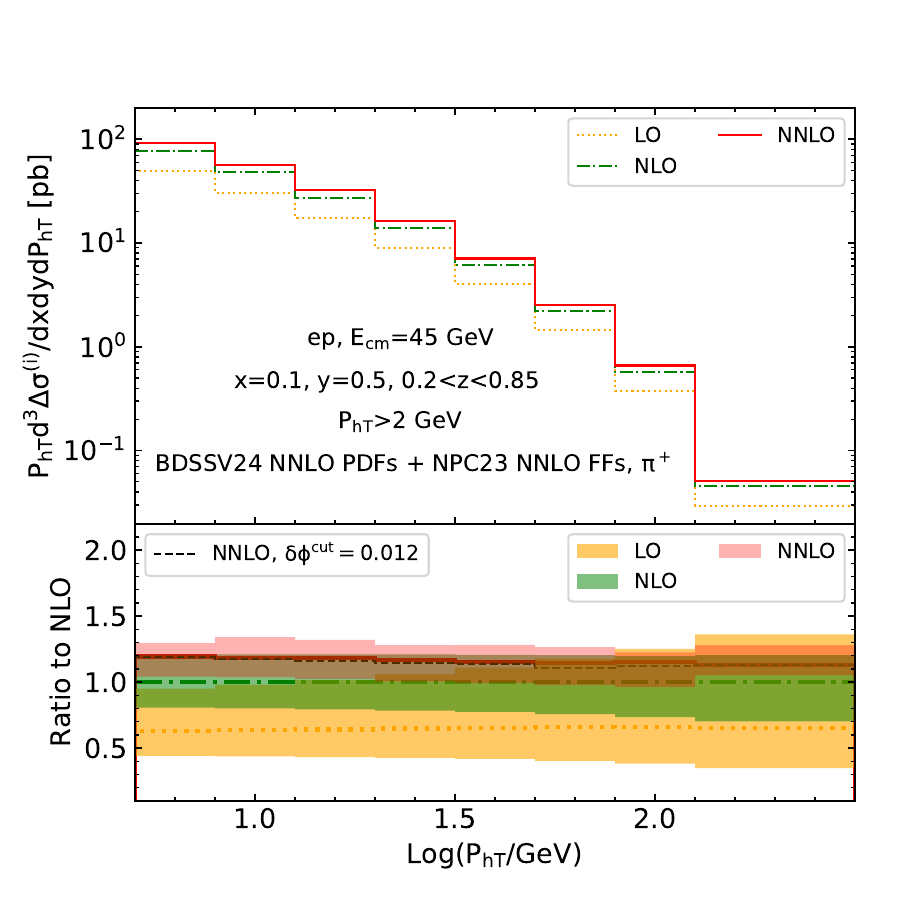}
    \vspace{-0.5cm}
    \caption{%
     Transverse-momentum spectrum for single-inclusive $\pi^+$ production in polarized $ep$ collisions at the center-of-mass energy $E_{\rm cm}=45$~GeV, with kinematics fixed at $x=0.1$ and $y=0.5$, and $0.2 < z < 0.85$. 
     The default cutoff of $\delta\phi^{\rm cut}$ is set to 0.02.
     The orange dotted, green dot-dashed, and red solid curves represent the LO, NLO and NNLO predictions, respectively. The lower panel displays the ratio of these distributions to the central NLO result, with bands indicating the scale uncertainties.
     The black dashed line denotes the NNLO result calculated with $\delta\phi^{\rm cut}=0.012$.
    }
    \label{fig:phT_dist}
\end{figure}

\paragraph*{Numerical results.---}

\begin{figure}[t]
    \includegraphics[width=0.95\linewidth]{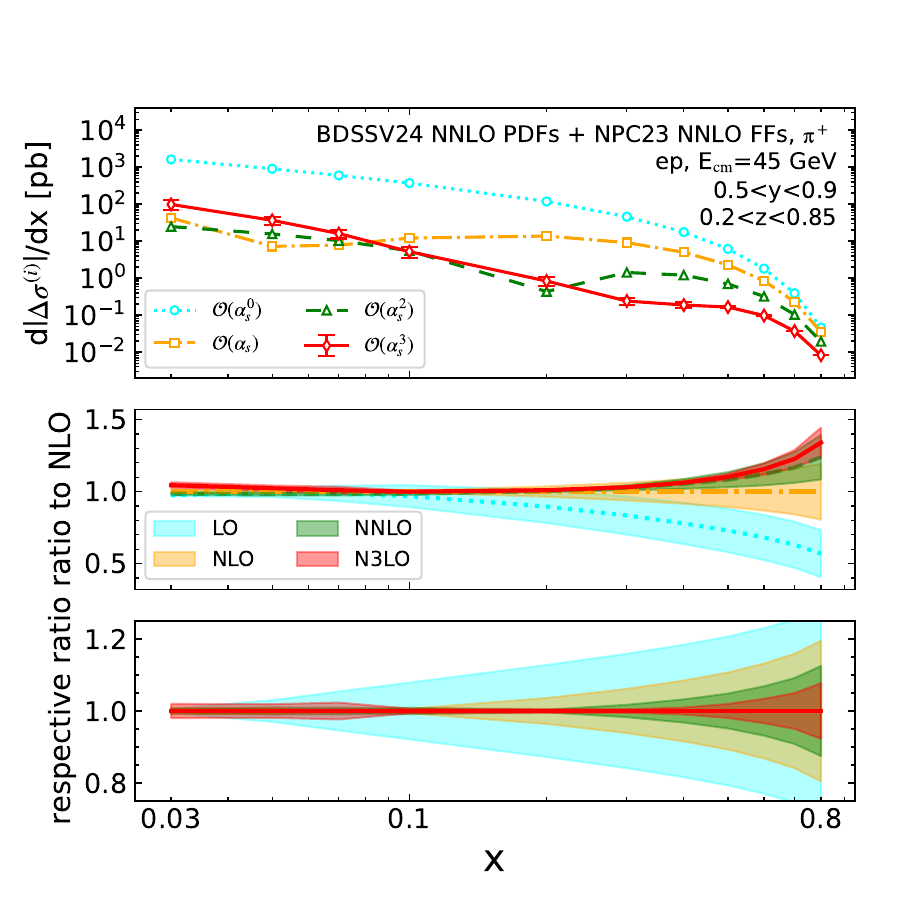}
    \vspace{-0.5cm}
    \caption{%
    Spin-dependent cross section as a function of the Bjorken variable $x$, for inclusive $\pi^+$ production in polarized $ep$ collisions at the center-of-mass energy $E_{\rm cm}=45$~GeV, with $0.5 < y < 0.9$, and $0.2 < z < 0.85$. 
    In all panels, LO, NLO, NNLO, and N$^3$LO corrections or predictions are represented by cyan, orange, green, and red curves, respectively. 
    In the upper panel, the error bars of N$^3$LO corrections indicate the Monte Carlo statistical uncertainties. 
    The middle (lower) panel presents the ratio of each prediction to the central prediction at NLO (the respective order). 
    The colored bands indicate the associated scale uncertainties.
    }
    \label{fig:x_dist}
\end{figure}

\begin{figure}[t]
    \includegraphics[width=0.95\linewidth]{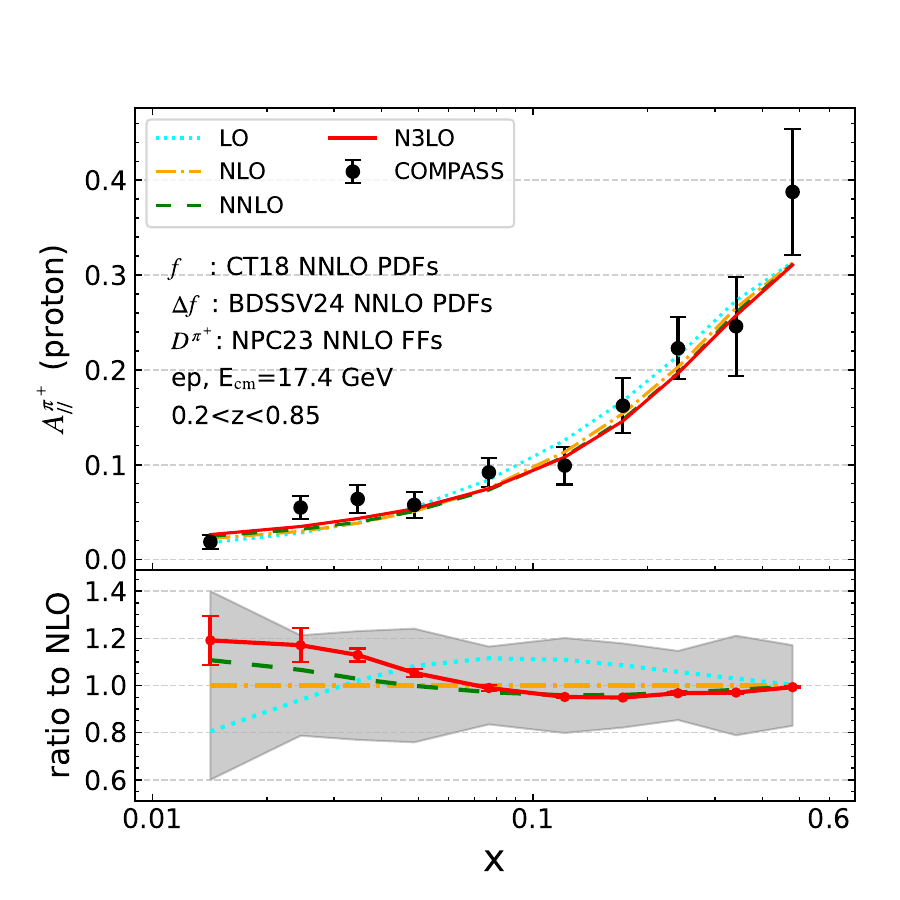}
    \vspace{-0.5cm}
    \caption{%
    Double-spin asymmetry as a function of $x$ compared with the COMPASS data (black points) for polarized $ep$ collisions at the center-of-mass energy $E_{\rm cm}=17.4$~GeV, with $0.2 < z < 0.85$. 
    The LO, NLO, NNLO, and N$^3$LO predictions are shown by the cyan, orange, green, and red curves, respectively. 
    The lower panel shows the ratio of the theoretical predictions to the central NLO predictions, and the gray band indicates the total experimental uncertainty.
    }
    \label{fig:x_dist_data}
\end{figure}

In this section we present phenomenological studies of polarized SIDIS at N$^3$LO in QCD.
We consider polarized electron-proton collisions at the center-of-mass energy of the EIC ($\sqrt s=45$ GeV) and use the BDSSV24 polarized PDF set at NNLO~\cite{Borsa:2024mss} with 5 active quark flavors and the strong coupling constant $\alpha_s(M_Z)=0.118$.  
We calculate differential cross sections for the charged pion ($\pi^+$) production with NPC23 NNLO FFs~\cite{Gao:2025hlm} through all orders, and use a nominal choice of $\mu=Q$ for both the renormalization and factorization scales. 
We assess scale uncertainties by varying $\mu$ by a factor of two around the central value.
For simplicity, we have neglected contributions from the $Z$ boson exchange, which are small for the energies considered.
We first present the spin-dependent cross section for hadron production at finite transverse momentum $P_{hT}$ calculated up to $\mathcal O(\alpha_s^3)$ (NNLO). 
In this case, only a single cutoff is needed which we set as $\delta\phi^{\rm cut}=0.02$.
Fig.~\ref{fig:phT_dist} displays the resulting $P_{hT}$ spectrum for polarized $ep$ collisions, tailored to the kinematics of the future EIC including $x=0.1$, $y=0.5$ and $0.2<z<0.85$.
In the upper panel, the LO, NLO, and NNLO predictions are denoted by orange dotted, green dot-dashed, and red solid curves, respectively. 
The lower panel presents the ratio to the central NLO result, with shaded bands indicating the scale uncertainties. 
We observe that NLO corrections enhance the cross section by approximately $60\%$. 
Relative to NLO, the NNLO corrections yield a further increase of about $20\%$ at $P_{\rm hT} \sim 2$~GeV, which steadily decreases to $13\%$ at $P_{\rm hT} \sim 10$~GeV.
Crucially, the scale uncertainties systematically diminish at each successive perturbative order. 
To validate the numerical result, we include an alternative NNLO calculation using a different phase-space cutoff of $\delta\phi^{\rm cut}=0.012$ (black dashed line). 
The precise agreement between the two results obtained with $\delta\phi^{\mathrm{cut}}=0.02$ and $0.012$ demonstrates the numerical stability of our framework and the residual power corrections being negligible.
Fig.~\ref{fig:cutoff_dep} presents a benchmark calculation of the $\mathcal{O}(\alpha_s^2)$ (NNLO) and $\mathcal{O}(\alpha_s^3)$ (N$^3$LO) corrections to the spin-dependent cross section of inclusive charged pion production $\d^2\Delta\sigma^{(i)}/\d x\d y$ at the EIC.
The kinematics are fixed at $x=0.6$ and $y=0.658$, with integration over $0.2<z<0.85$. 
The results are shown for six representative partonic channels, including $uu$, $ug$, $gu$, $gg$, $u\bar u$, and $u d$, labeled according to the flavors of the incoming and fragmenting partons.
Among them, the $uu$ channel is dominant due to both favored PDFs and FFs.
Note that we have multiplied the cross section of each channel by different factors for easy comparisons. 
We scan over the slicing parameter $\Delta$ while choosing two representative values of $\lambda$, 0.3 and 0.2 (0.2 and 0.1) at N$^3$LO (NNLO).
The NNLO results (red points) exhibit excellent flatness over $0.02 < \Delta< 0.1$, confirming a precise cancellation of infrared divergences, and converge to the reference values calculated in Refs.~\cite{Bonino:2024wgg, Goyal:2024tmo}. 
At N$^3$LO (blue points), the results similarly converge to a constant plateau.
Results from two different choices of $\lambda$ are consistent for small values of $\lambda$.
The solid cyan curves denote asymptotic fits incorporating power corrections~\cite{Grazzini:2017mhc, Ebert:2019zkb} for the $\mathcal{O}(\alpha_s^3)$ data, with a three-parameter functional form $\Delta\sigma^{(3)}=\Delta^2(a_5\ln^n(\Delta/4)+a_4\ln^{n-1}(\Delta/4))+b_0$ where $n=5$ for the $uu$ channel, $n=4$ for the $ug$, $gu$ channels and $n=3$ for the $gg$, $ud$ and $u\bar{u}$ channels.
The red lines are connections of the $\mathcal{O}({\alpha_s^2})$ data points. 
From the fit we decide that a choice of $\Delta=0.1$ in the N$^3$LO calculations is sufficient for keeping power corrections under control in a phenomenological study.
Using even smaller values of $\Delta$ would require extensive computational resources in order to maintain comparable numerical precision.  
The stability observed across both quark- and gluon-initiated channels establishes the robustness of our method.
Overall we find good convergence of the perturbative expansion with N$^3$LO corrections smaller than NNLO corrections, especially for the favored $uu$ channel.
For the $ud$, $u\bar u$ and $gg$ channels, which first appear at NNLO, the N$^3$LO corrections can be comparable to or even larger than the NNLO corrections. 
Fig.~\ref{fig:x_dist} illustrates the Bjorken-$x$ dependence of the spin-dependent cross section under the aforementioned EIC kinematics, integrating over the range $0.2 < z < 0.85$ and $0.5 < y < 0.9$. 
The upper panel presents the LO prediction of $\d\Delta\sigma/\d x$, alongside the absolute corrections at different perturbative orders.
The middle (lower) panel presents the ratio of the full prediction at successive orders to the central prediction at NLO (the respective order), with shaded bands indicating the scale uncertainties. 
The relative size of NNLO corrections exhibits a distinct behavior of being large for $x\gtrsim 0.3$, reaching its minimum at $x\sim 0.2$, and rising again for $x<0.1$.
Similarly, for the large-$x$ region, the N$^3$LO corrections can be as large as 10\% but are significantly smaller than the NNLO corrections.
Despite the large magnitude of these corrections, the inclusion of the higher-order corrections drastically reduces the scale uncertainties in the large-$x$ region.
Such corrections are potentially dominated by the threshold logarithms which are known to higher orders~\cite{Abele:2022wuy, Goyal:2025bzf}.
In the intermediate-$x$ region, both the NNLO and N$^3$LO corrections are below 1\%, and the scale variations are negligible. 
Lastly, in the low-$x$ region, the perturbative convergence is much worse, possibly due to small-$x$ logarithmic corrections.
The NNLO corrections can be comparable to or larger than the NLO corrections.
The N$^3$LO corrections are even larger, at the level of a few percent.
The residual scale variations are slightly increased to about 2\%.
We are aware that in the low-$x$ region our numerical calculations converge much more slowly, leading to MC uncertainties of more than 1\% of the LO predictions.
Fig.~\ref{fig:x_dist_data} presents predictions on the double-spin asymmetry, defined as the ratio of spin-dependent and unpolarized cross sections, $A_{//}^{\pi^+}(x) = \Delta\sigma^{\pi^+}(x)/\sigma^{\pi^+}(x)$, for inclusive $\pi^+$ production.
Calculations on the unpolarized cross section $\sigma^{\pi^+}(x)$ up to N$^3$LO are detailed in Ref.~\cite{Dong:2026uvw} with the CT18 NNLO PDFs~\cite{Hou:2019efy}.
The predictions are compared to the COMPASS measurement with a polarized proton target at the center-of-mass energy of $E_{\rm cm}=17.4$~GeV, and integrated over $0.2 < z < 0.85$~\cite{COMPASS:2010hwr}.
The experimental results are presented in the form of
$A_1^{\pi^+}$ as the ratio of the polarized and unpolarized structure functions. 
We convert the data back to $A_{//}$ through the relation
\begin{align}
    A_{//}^{\pi^+} \simeq D\, A_1^{\pi^+}\,,
\end{align}
where $D$ is the depolarization factor, whose $x$-dependent values are taken from Ref.~\cite{Wilfert:2017mvt}.
The theoretical predictions are evaluated at the effective $x$ and $Q^2$ values provided by the experimental group, with the latter being as low as 2 GeV$^2$.
Both the spin-dependent and unpolarized cross sections are calculated to the prescribed order.
From the lower panel of Fig.~\ref{fig:x_dist_data}, we observe excellent perturbative convergence of the theoretical predictions for $x\gtrsim 0.08$, with N$^3$LO predictions almost indistinguishable from the NNLO ones. 
Both the NLO and NNLO corrections exhibit a turning point at $x$ around 0.4.  
For $x<0.05$ the N$^3$LO corrections are sizeable, indicating poor perturbative convergence at small-$x$ and low-$Q^2$.
The Monte Carlo uncertainties, represented by the red error bars, are also significant due to slow convergence of numerical calculations for both spin-dependent and unpolarized cross sections at small-$x$.
The half-width of the shaded band indicates the total uncertainties of current COMPASS measurements which are much larger than the higher-order QCD corrections in general.
As a result, the NLO, NNLO and N$^3$LO predictions describe the data equally well.

\paragraph*{Summary and Outlook.---}
In summary, we have presented the first N$^3$LO QCD predictions for polarized SIDIS in a fully differential form, together with NNLO predictions for the hadron transverse-momentum spectrum. These results are obtained by extending the two-dimensional transverse-momentum subtraction framework to polarized scattering.
The N$^3$LO corrections can reach about $10\%$ at large-$x$ and substantially reduce the perturbative scale uncertainties over most of the kinematic region. 
Together with previous unpolarized calculations, we obtain consistent N$^3$LO predictions for the double-spin asymmetry and compare them with COMPASS measurements.
The perturbative series converges well for $x\gtrsim0.08$, whereas the $x<0.05$ bins at low $Q^2$ show poorer convergence and larger Monte Carlo integration uncertainties.
Our results provide an important step forward for high-precision determinations of helicity PDFs and TMDs at the future EIC.

\paragraph*{Acknowledgments.---}
The work of L.~D. and J.~G. is supported by the National Natural Science Foundation of China (NSFC) under Grant No.~12275173 and the Shanghai Municipal Education Commission under Grant No.~2024AIZD007.
S.~F. and D.~Y.~S. are supported by the National Natural Science Foundation of China under Grant No.~12275052, No.~12147101, No.~12547102, and the Innovation Program for Quantum Science and Technology under Grant No. 2024ZD0300101.
H.~T.~L. is supported by the National Natural Science Foundation of China under Grant No.~12275156, No.~12321005, and the Shandong Provincial Department of Science and Technology under Project No.~TQ012025001.
B.~Z. is supported by the Hainan University Startup Fund under Grant No.~XJ2600002569.
The computations in this paper were run on the Siyuan-1 cluster supported by the Center for High Performance Computing at Shanghai Jiao Tong University.

\bibliographystyle{apsrev4-2}
\bibliography{psidis}

\end{document}